 \documentstyle[epsf]{mn}

\newif\ifAMStwofonts

\ifoldfss
  \ifCUPmtlplainloaded \else
    \NewTextAlphabet{textbfit} {cmbxti10} {}
    \NewTextAlphabet{textbfss} {cmssbx10} {}
    \NewMathAlphabet{mathbfit} {cmbxti10} {} 
    \NewMathAlphabet{mathbfss} {cmssbx10} {} 
  \fi
  \ifAMStwofonts
    \ifCUPmtlplainloaded \else
      \NewSymbolFont{upmath} {eurm10}
      \NewSymbolFont{AMSa} {msam10}
      \NewMathSymbol{\upi}     {0}{upmath}{19}
      \NewMathSymbol{\umu}     {0}{upmath}{16}
      \NewMathSymbol{\upartial}{0}{upmath}{40}
      \NewMathSymbol{\leqslant}{3}{AMSa}{36}
      \NewMathSymbol{\geqslant}{3}{AMSa}{3E}

      \let\leq=\leqslant 
      \let\geq=\geqslant 
    \fi
  \fi
\fi 

\ifnfssone
  \newmathalphabet{\mathit}
  \addtoversion{normal}{\mathit}{cmr}{m}{it}
  \addtoversion{bold}{\mathit}{cmr}{bx}{it}
  \newmathalphabet{\mathbfit} 
  \addtoversion{normal}{\mathbfit}{cmr}{bx}{it}
  \addtoversion{bold}{\mathbfit}{cmr}{bx}{it}
  \newmathalphabet{\mathbfss} 
  \addtoversion{normal}{\mathbfss}{cmss}{bx}{n}
  \addtoversion{bold}{\mathbfss}{cmss}{bx}{n}
  \ifAMStwofonts
    \ifCUPmtlplainloaded \else
      \UseAMStwoboldmath
      \makeatletter
      \new@mathgroup\upmath@group
      \define@mathgroup\mv@normal\upmath@group{eur}{m}{n}
      \define@mathgroup\mv@bold\upmath@group{eur}{b}{n}
      \edef\UPM{\hexnumber\upmath@group}
      \new@mathgroup\amsa@group
      \define@mathgroup\mv@normal\amsa@group{msa}{m}{n}
      \define@mathgroup\mv@bold\amsa@group{msa}{m}{n}
      \edef\AMSa{\hexnumber\amsa@group}
      \makeatother
      \mathchardef\upi="0\UPM19
      \mathchardef\umu="0\UPM16
      \mathchardef\upartial="0\UPM40
      \mathchardef\leqslant="3\AMSa36
      \mathchardef\geqslant="3\AMSa3E

      \let\leq=\leqslant 
      \let\geq=\geqslant 
    \fi
  \fi
\fi 

\ifnfsstwo
  \DeclareMathAlphabet{\mathbfit}{OT1}{cmr}{bx}{it}
  \SetMathAlphabet\mathbfit{bold}{OT1}{cmr}{bx}{it}
  \DeclareMathAlphabet{\mathbfss}{OT1}{cmss}{bx}{n}
  \SetMathAlphabet\mathbfss{bold}{OT1}{cmss}{bx}{n}
  \ifAMStwofonts
    \ifCUPmtlplainloaded \else
      \DeclareSymbolFont{UPM}{U}{eur}{m}{n}
      \SetSymbolFont{UPM}{bold}{U}{eur}{b}{n}
      \DeclareSymbolFont{AMSa}{U}{msa}{m}{n}
      \DeclareMathSymbol{\upi}{0}{UPM}{"19}
      \DeclareMathSymbol{\umu}{0}{UPM}{"16}
      \DeclareMathSymbol{\upartial}{0}{UPM}{"40}
      \DeclareMathSymbol{\leqslant}{3}{AMSa}{"36}
      \DeclareMathSymbol{\geqslant}{3}{AMSa}{"3E}

      \let\leq=\leqslant 
      \let\geq=\geqslant 
    \fi
  \fi
\fi 

\ifCUPmtlplainloaded \else
  \ifAMStwofonts \else 
    \def\upi{\pi}
    \def\umu{\mu}
    \def\upartial{\partial}
  \fi
\fi

\newcommand{\Nco}{N_{\rm e}^*}		

\title{Composition of active galactic nuclei jets:
       pair-plasma dominance in the 3C~345 and 3C~279 jets}
\author[Kouichi Hirotani]
       {Kouichi Hirotani \\
        National Astronomical Observatory,\\
	Osawa, Mitaka, Tokyo 181-8588, Japan\\
        hirotani@hotaka.mtk.nao.ac.jp
       }
\date{Accepted .
      Received 2000 September;
      in original form 2000 September}

\pagerange{\pageref{firstpage}--\pageref{lastpage}}
\pubyear{1997}

\begin{document}

\maketitle

\label{firstpage}

\begin{abstract}
We investigate whether the parsec-scale jets of 
quasars 3C~345 and 3C~279 are dominated by 
a normal (proton-electron) plasma 
or a pair (electron-positron) plasma.
We first present a new method to compute the kinetic luminosity
of a conical jet by using the core size observed at a single
very long baseline interferometry frequency.
The deduced kinetic luminosity gives electron densities
of individual radio-emitting components
as a function of the composition.
We next constrain the electron density independently
by using the theory of synchrotron self-absorption.
Comparing the two densities, we can discriminate the composition.
We then apply this procedure to the five components in the 3C~345 jet
and find that they are pair-plasma dominated at 14 epochs out of 
the total 19 epochs at which the turnover frequencies are reported, 
provided that the bulk Lorentz factor is less than 15
throughout the jet.
We also investigate the composition of the 3C~279 jet 
and demonstrate that its two components are likely pair-plasma dominated 
at all the four epochs,
provided that their Doppler factors are less than 10,
which are consistent with observations.
The conclusions do not depend on the lower cutoff energy of 
radiating particles.
\end{abstract}


\section{Introduction}

The study of extragalactic jets on parsec scales is 
astrophysically interesting in the context of 
the activities of the central engines of 
active galactic nuclei (AGN).
In particular, a determination of their matter content 
would be an important step in the study of jet formation, 
propagation and emission.
There are two main candidates for their matter content:
a \lq normal plasma' consisting of
protons and relativistic electrons
(for numerical simulations of shock fronts in a 
very long baseline interferometry jet, see e.g.,
 G$\acute{\rm o}$mez et al. 1993, 1994a,b),
and a \lq pair plasma' consisting only of relativistic electrons
and positrons
(for theoretical studies of two-fluid concept, see
 Sol, Pelletier \& Ass$\acute{\rm e}$o 1989;
 Despringre \& Fraix-Burnet 1997).
Distinguishing between these possibilities is crucial for
understanding the physical processes occurring close to the 
central engine (presumably a supermassive black hole)
in the nucleus.

Very long baseline interferometry (VLBI) 
is uniquely suited to the study of the matter content of 
pc-scale jets,
because other observational techniques cannot image 
at milliarcsecond resolution and must resort to indirect means of
studying the active nucleus.
Recently, Reynolds et al. (1996) analyzed historical VLBI data
of the M87 jet at 5 GHz (Pauliny-Toth et al. 1981)
and concluded that the core is probably dominated 
by an $e^\pm$ plasma.
In the analysis, they utilized the standard theory of synchrotron 
self absorption (SSA) to constrain the magnetic field, $B$ [G], and 
the proper number density of electrons, 
$\Nco$ [1/cm${}^3$], 
and derived the following condition
for the core to be optically thick for SSA:
$ \Nco B^2 > 2 \delta_{\rm max}^{-2} $,
where $\delta_{\max}$ refers to the upper limit of the 
Doppler factor of the fluid's bulk motion.
This condition is, however, applicable only for the VLBI observations
of M87 core at epochs September 1972 and March 1973.
Therefore, in order to apply the analogous method to other AGN jets
or to the M87 jet at other epochs, 
we must derive a more general condition.

On these grounds, Hirotani et al. (1999, hereafter Paper I) 
generalized the condition
$ \Nco B^2 > 2 \delta_{\rm max}^{-2} $
and applied it to the 3C~279 jet on parsec scales. 
In that paper, they revealed that the core and components C3 and C4,
of which spectra are reported,
are dominated by a pair plasma.
It is interesting to note that the same conclusion 
that component CW in the 3C~279 jet is dominated by a pair plasma
was derived by an independent method by Wardle et al. (1998),
who studied the circularly and linearly
polarized radio emission from 3C~279.

Subsequently, Hirotani et al. (2000, hereafter Paper II)
applied the same method to the 3C~345 jet.
Deducing the kinetic luminosity from a reported core-position
offset (Lobanov 1998), 
they demonstrated that components C2, C3, and C4 at epoch 1982.0,
C5 at 1990.55, and C7 at four epochs are dominated by a pair plasma.

In the present paper,
we revisit the two blazers,
adding spectral information at 2 epochs for 3C~279 and 11 epochs
for 3C~345.
In the next section, we first improve the method presented in Paper II
and demonstrate that the kinetic luminosity, $L_{\rm kin}$,
can be inferred from the VLBI core size at a single frequency,
rather than from the core-position offset,
which requires more than two VLBI frequencies.
We then provide the method to deduce the electron density, 
$\Nco({\rm nml})$, from $L_{\rm kin}$,
assuming a normal-plasma dominance.
In~\S 3, we describe an independent method to deduce the electron density,
$\Nco({\rm SSA})$, by utilizing the theory of SSA;
this method is based on the generalized condition of
$\Nco B^2 > 2 \delta_{\rm max}{}^{-2}$ in Reynolds et al. (1996).
If $\Nco({\rm nml}) < \Nco({\rm SSA})$ holds for a radio-emitting
component, we can rule out the possibility of a normal-plasma dominance.
In \S~4, we apply the method to the 3C~345 jet and find
that components C2, C3, C4, C5, and C7 are dominated by a pair plasma
as a whole.
We next investigate the 3C~279 jet in \S~5 
and demonstrate that components C3 and C4
are pair-dominated,
provided that their Doppler factors are less than 10.
In the final section, 
we discuss the possibility of the entrainment of the ambient matter
as a jet propagates downstream.

We use a Hubble constant $H_0 = 65 h$ km/s/Mpc and $q_0 =0.5$
throughout this paper.
Spectral index $\alpha$ is defined 
such that $S_\nu \propto \nu^\alpha$.

\section{Kinetic Luminosity Deduced From the Core Size}  
\label{sec:Lkin_Core_Size}

In this section, we present a new method to deduce the
kinetic luminosity of a conical jet from the core size observed
at a single VLBI frequency in \S\S~\ref{sec:conical}--\ref{sec:opening}.
We then describe how to compute the electron density 
from the kinetic luminosity in \S~\ref{sec:Ne_kin}.

\subsection{Conical Jet Geometry}
\label{sec:conical}

\begin{figure} 
\centerline{ \epsfxsize=8cm \epsfbox[100 200 400 400]{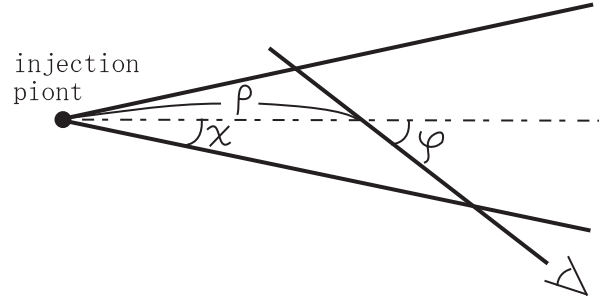} } 
\caption{\label{fig:conical} 
Schematic figure of a conical jet in the core region 
with half opening angle $\chi$ in the observer's frame.
        }
\end{figure}

We assume that the parsec-scale jet close to the central engine
propagates conically with a half opening angle $\chi$ 
(fig.~\ref{fig:conical}).
Then the optical depth $\tau$ for 
synchrotron self absorption at distance $\rho$ from the injection point,
is given by
\begin{equation}
  \tau_{\nu}(\rho) 
  = \left[ \frac{\rho \sin\chi}{\sin(\varphi +\chi)}
          +\frac{\rho \sin\chi}{\sin(\varphi -\chi)}
    \right]
    \alpha_{\nu}, 
\label{eq:tau1}
\end{equation}
where $\varphi$ is the viewing angle
and $\alpha_\nu$ [1/cm] refers to the effective absorption 
(i.e., absorption minus stimulated emission) coefficient.
For a small half opening angle ($\chi \ll 1$),
this equation can be approximated as
\begin{equation}
  \tau_{\nu}(\rho)= 2\rho \frac{\chi}{\sin\varphi} \alpha_{\nu}
  \label{eq:tau2}
\end{equation}
Note that $\chi/\sin\varphi$ denotes the projected, observed
opening angle.

Since $\tau_\nu$ and $\rho \chi$ are Lorentz invariants, we obtain
\begin{equation}
  \frac{\alpha_\nu}{\sin\varphi}
  = \frac{\alpha_\nu^*}{\sin\varphi^*},
  \label{eq:LI-1}
\end{equation}
where a quantity with an asterisk is measured in the co-moving frame,
while that without an asterisk in the observer's frame.
Since $\nu \alpha_\nu$ is also Lorentz invariant, 
equation (\ref{eq:LI-1}) gives
\begin{equation}
  \frac{\sin\varphi^*}{\sin\varphi}
  = \frac{\nu}{\nu^*}
  = \frac{\delta}{1+z} .
  \label{eq:LI-3}
\end{equation}
Combining equations (\ref{eq:tau2}) and (\ref{eq:LI-3}), we obtain
\begin{equation}
  \tau_\nu
      = \frac{1+z}{\delta} \frac{2\rho\chi}{\sin\varphi} \alpha_\nu^* 
      = \frac{1+z}{\delta} \frac{\theta_{\rm d}}{\sin\varphi}
        \frac{D_{\rm L}}{(1+z)^2} \alpha_\nu^* ,
  \label{eq:o-depth-2}
\end{equation}
where $D_{\rm L}/(1+z)^2$ is the angular diameter distance to the AGN,
and $\theta_{\rm d}$ is the angular diameter of the component
in the perpendicular direction of the jet propagation in radian units;
the Doppler factor, $\delta$, is defined by
\begin{equation}
  \delta \equiv \frac{1}{\Gamma(1-\beta\cos\varphi_{\rm A})},
  \label{eq:def_delta}
\end{equation}
where $\Gamma \equiv 1 / \sqrt{1-\beta^2}$ is the bulk Lorentz factor
of the jet component moving with velocity $\beta c$,
and $\varphi_{\rm A}$ refers to the viewing angle in the
AGN-rest frame.
It is related with $\varphi$, viewing angle in the observer's frame, as
\begin{equation}
  \cos\varphi_{\rm A}
    = \frac{\cos\varphi -\beta_{\rm AGN}}
           {1-\beta_{\rm AGN}\cos\varphi},
  \label{eq:angles} 
\end{equation}
where $\beta_{\rm AGN}$ is defined by
\begin{equation}
  \sqrt{\frac{1+\beta_{\rm AGN}}{1-\beta_{\rm AGN}}} = \frac{1}{1+z}.
  \label{eq:def_betaAGN}
\end{equation}
For 3C~345($z= 0.595$) for instance, 
if a photon is ejected with $\varphi_{\rm A}=2^\circ$ to the observer
in the AGN-rest frame,
we observe it propagates in the direction $\varphi= 3.2^\circ$
due to the redshift ($\beta_{\rm AGN}<0$).
We should distinguish $\varphi$ and $\varphi_{\rm A}$, 
because $\delta$, which appears in equation~(\ref{eq:LI-3}),
should be defined in the AGN-rest frame.

We assume that the electron number density between
energies $\gamma m_{\rm e}c^2$ and $(\gamma+d\gamma) m_{\rm e}c^2$ 
is expressed by a power law as
\begin{equation}
  \frac{dN_{\rm e}^*}{d\gamma} = k_{\rm e}^\ast \gamma^{2\alpha-1}
  \qquad (\gamma_{\rm min} < \gamma < \gamma_{\rm max}).
  \label{eq:ele-dist}
\end{equation}
Integrating $d\Nco/d\gamma$ from $\gamma_{\rm min}$ to 
$\gamma_{\rm max}$,
and assuming $\gamma_{\rm max} \gg \gamma_{\rm min}$
and $\alpha<0$,
we obtain the electron number density
\begin{equation}
  \Nco = \frac{\gamma_{\rm min}{}^{2\alpha}}{-2\alpha} 
         k_{\rm e}^\ast.
  \label{eq:Ne-ke}
\end{equation}
Moreover, averaging over pitch angles of the isotropic electron power-law
distribution (eq. [\ref{eq:ele-dist}]), 
we can write down $\alpha_\nu$ in the co-moving frame of the jet as
(Le Roux 1961, Ginzburg \& Syrovatskii 1965)
\begin{equation}
  \alpha_\nu^* = C(\alpha) r_\circ{}^2 k_{\rm e}^\ast
                 \frac{\nu_\circ}{\nu^*}
		 \left( \frac{\nu_{\rm B}}{\nu^*} 
                 \right)^{(-2\alpha+3)/2} ,
  \label{eq:abs-coeff}
\end{equation}
where $\nu_\circ \equiv c/r_\circ \equiv c/[e^2 / (m_{\rm e} c^2)]
= 1.063 \times 10^{14}$ GHz
and $\nu_{\rm B} \equiv eB / (2\pi m_{\rm e}c)$;
$e$, $m_{\rm e}$, and $c$ refer to the charge on an electron,
the rest mass of an electron, and the speed of light, respectively.
The coefficient $C(\alpha)$ is given in Table~1 of Gould (1979)
and also in table~1 in this paper.

\subsection{Scaling Law}
\label{sec:scaling}

Let us introduce a dimensionless variable $r \equiv \rho/r_1$
to measure $\rho$ in parsec units, where $r_1=1$ pc.
Assuming that there is neither particle production nor annihilation,
we obtain the following scaling law for a conical jet
from the conservation of the particles: 
\begin{eqnarray}
  \Nco = N_1 r^{-n},
  \label{eq:scaling_N}
\end{eqnarray}
where $n=2$; $N_1$ refers to the value of 
$\Nco$ at $r=1$ (i.e., at 1 pc from the core in the observer's frame).
In the same manner, the magnetic flux conservation 
in a stationary, supersonic magnetohydrodynamic jet leads to
\begin{eqnarray}
  B= B_1 r^{-m},
  \label{eq:scaling_B}
\end{eqnarray}
where $m=1$; $B_1$ refers to the values of $B$ at $r=1$.
We further introduce the following dimensionless variables:
\begin{eqnarray}
  x_{\rm N} \equiv r_1 r_\circ{}^2 N_1,
  \nonumber \\
  x_{\rm B} \equiv \nu_{\rm B_1}/\nu_\circ
    = \frac{eB_1}{2\pi m_{\rm e}c}.
  \label{eq:def_x}
\end{eqnarray}
Utilizing equation (\ref{eq:abs-coeff}), 
and rewriting $k_{\rm e}^\ast$ and $\nu_{\rm B}$ in terms of 
$x_{\rm N}$ and $x_{\rm B}$,
we obtain from the left equality in equation (\ref{eq:o-depth-2})
\begin{eqnarray}
  \tau_\nu 
     &=& C(\alpha) \frac{2\chi}{\sin\varphi}
             \frac{-2\alpha}{\gamma_{\rm min}{}^{2\alpha}}
             \left(\frac{1+z}{\delta}\right)^{-\epsilon}
             \left(\frac{\nu}{\nu_\circ}\right)^{-1-\epsilon}
             r^{1-n-m\epsilon}
     \nonumber\\
     & & \times \quad
             x_{\rm N} x_{\rm B}{}^\epsilon,
  \label{eq:tau4}
\end{eqnarray}
where $\epsilon \equiv 3/2 -\alpha$.

At a given frequency $\nu$, 
the flux density will peak at the position where $\tau_\nu$ becomes unity.
Thus setting $\tau=1$ and solving equation (\ref{eq:tau4}) 
for $r$, we obtain the distance from the VLBI core 
observed at frequency $\nu$ from the injection point 
(or the central engine) as (Lobanov 1998)
\begin{equation}
  r(\nu) = \left( x_{\rm B}{}^{k_{\rm b}} F \frac{\nu_\circ}{\nu}
           \right)^{1/k_{\rm r}}
  \label{eq:core_rad} 
\end{equation}
where 
\begin{equation}
  f(\alpha) \equiv \left[ C(\alpha) \frac{2\chi}{\sin\varphi}
                          \frac{-2\alpha}{\gamma_{\rm min}{}^{2\alpha}}
                          \left(\frac{\delta}{1+z} \right)^{\epsilon} 
                          x_{\rm N} 
                   \right]^{1/(\epsilon+1)}
\end{equation}
\begin{equation}
  k_{\rm b} \equiv \frac{3-2\alpha}{5-2\alpha},
\end{equation}
\begin{equation}
  k_{\rm r} \equiv \frac{(3-2\alpha)m+2n-2}{5-2\alpha}. 
\end{equation}

\subsection{Core-Position Offset}
\label{sec:CPO}

If we mesure $r(\nu)$ at  two different frequencies
(say $\nu_{\rm a}$ and $\nu_{\rm b}$),
equation (\ref{eq:core_rad}) gives the dimensionless, projected 
distance of
$r(\nu_{\rm a})-r(\nu_{\rm b})$ as 
\begin{eqnarray}
  \Delta r_{\rm proj} 
  &=& \left[ r(\nu_{\rm a}) - r(\nu_{\rm b}) \right] \sin\varphi
  \nonumber\\
  &=& (x_{\rm B}{}^{k_{\rm b}} f \nu_\circ)^{1/k_{\rm r}}
      \frac{\nu_{\rm b}^{1/k_{\rm r}} - \nu_{\rm a}^{1/k_{\rm r}}}
      {\nu_{\rm a}^{1/k_{\rm r}} \nu_{\rm b}^{1/k_{\rm r}}}
      \sin\varphi.
  \label{eq:del_jet}
\end{eqnarray}
Here, $\Delta r_{\rm proj}$ is in pc units;
therefore, $r_1 \Delta r_{\rm proj}$ represents the projected
distance of the two VLBI cores (in cm, say).
That is, $r_1 \Delta r_{\rm proj}$ equals
$4.85 \cdot 10^{-9} \Delta r_{\rm mas} D_{\rm L}/(1+z)^2$
in equation~(4) in Lobanov (1998),
where $\Delta r_{\rm mas}$ refers to the
core-position difference in mas.
Defining the core-position offset as (Lobanov 1998)
\begin{equation}
  \Omega_{r \nu} \equiv 
    r_1 \Delta r_{\rm proj}
    \frac{\nu_{\rm a}^{1/k_{\rm r}} \nu_{\rm b}^{1/k_{\rm r}}}
         {\nu_{\rm b}^{1/k_{\rm r}} - \nu_{\rm a}^{1/k_{\rm r}}},
  \label{eq:def_CPO}
\end{equation}
we obtain
\begin{equation}
  \frac{\Omega_{r\nu}}{r_1}
  = (x_{\rm B}^{k_{\rm b}} f \nu_\circ )^{1/k_{\rm r}} \sin\varphi
  \label{eq:CPO}
\end{equation}

To express $x_{\rm B}$ in terms of $x_{\rm N}$ and 
$\Omega_{r\nu}$, we solve equation (\ref{eq:CPO}) 
for $x_{\rm B}$ to obtain
\begin{equation}
  x_{\rm B} = \left( \frac{\Omega_{r\nu}}{r_1 \sin\varphi}
              \right)^{k_{\rm r}/k_{\rm b}}
              (f \nu_\circ)^{-1/k_{\rm b}}.
  \label{eq:xB}
\end{equation}
Note that $x_{\rm N}$ is included in $f=f(\alpha)$.

Setting $\nu_{\rm b} \rightarrow \infty$ in equation (\ref{eq:del_jet}), 
we obtain the absolute distance of the VLBI core
measured at $\nu$ from the central engine as (Lobanov 1998)
\begin{equation}
  r_{\rm core} (\nu) = \frac{\Omega_{r \nu}}{r_1 \sin \varphi}
                       \nu^{-1/k_{\rm r}}.
  \label{eq:Rcore}
\end{equation}
That is, once $\Omega_{r\nu}$ is obtained from multi-frequency VLBI
observations,
we can deduce the distance of the synchrotron-self-absorbing VLBI core 
from the central engine,
assuming the scaling laws of $\Nco$ and $B$ 
as equations (\ref{eq:scaling_N}) and (\ref{eq:scaling_B}).

We next represent $x_{\rm N}$ and $x_{\rm B}$ 
(or equivalently, $N_1$ and $B_1$) as a function of $\Omega_{r\nu}$.
To this end, we relate $\Nco$ and $B$ as follows:
\begin{equation}
  \Nco \gamma_{\min} m_{\rm e} c^2 = K \frac{B^2}{8 \pi}.
  \label{eq:equiP_0}
\end{equation}
When an energy equipartition between the radiating particles
and the magnetic field holds,
we obtain
\begin{equation}
  K = \frac{\gamma_{\rm min}}{\langle \gamma_- \rangle},
  \label{eq:K_1}
\end{equation}
where the averaged electron Lorentz factor, $\langle \gamma_- \rangle$
becomes
\begin{eqnarray}
  \langle \gamma_- \rangle
    &\equiv& 
        \frac{ \displaystyle{\int_{\gamma_{\rm min}}^{\gamma_{\rm max}}
               \gamma \cdot k_{\rm e}^\ast \gamma^{2\alpha-1} d\gamma }}
             { \displaystyle{\int_{\gamma_{\rm min}}^{\gamma_{\rm max}}
                            k_{\rm e}^\ast \gamma^{2\alpha-1} d\gamma }}
    \nonumber \\
    &=& \frac{2\alpha}{2\alpha+1} \gamma_{\rm min}
        \frac{(\gamma_{\rm max}/\gamma_{\rm min})^{2\alpha+1}-1}
             {(\gamma_{\rm max}/\gamma_{\rm min})^{2\alpha  }-1}.
  \label{eq:def_gamAVR}
\end{eqnarray}
for $\alpha<0$.
In the special case $\alpha \rightarrow -0.5$, it reduces to
\begin{equation}
  \langle \gamma_- \rangle 
    = \gamma_{\rm min} \ln\left(\frac{\gamma_{\rm max}}
                                     {\gamma_{\rm min}}
                          \right).
  \label{eq:def_gamAVR_2} 
\end{equation}

Substituting $N_{\rm e}^* = N_1 r^{-2}$ and $B = B_1 r^{-1}$ into 
equation~(\ref{eq:equiP_0}),
and replacing $N_1$ and $B_1$ with $x_{\rm N}$ and $x_{\rm B}$,
we obtain
\begin{equation}
  x_{\rm N}= \frac{\pi}{2} \frac{K}{\gamma_{\rm min}}
             \frac{r_1}{r_\circ}
             x_{\rm B}{}^2
  \label{eq:equiP_1}
\end{equation}
It is noteworthy that the assumptions of $n=2$ and $m=1$, 
which results in $k_{\rm r}=1$,
in consistent with the energy equipartition at an arbitrary distance, $r$.
Combining equations (\ref{eq:xB}) and (\ref{eq:equiP_1}),
we obtain
\begin{eqnarray}
  x_{\rm B}
  &=& \left( \frac{\Omega_{r\nu}/\nu_\circ}{r_1 \sin\varphi}
      \right)^{(5-2\alpha)/(7-2\alpha)}
  \nonumber \\
  && \hspace{-0.5 truecm} \times 
     \left[ \pi C(\alpha) \frac{\chi}{\sin\varphi}
             \frac{K}{\gamma_{\rm min}}
             \frac{r_1}{r_\circ}
             \frac{-2\alpha}{\gamma_{\rm min}{}^{2\alpha}}
             \left(\frac{\delta}{1+z}\right)^\epsilon
      \right]^{-2/(7-2\alpha)}.
  \label{eq:sol_xB}
\end{eqnarray}
For ordinary values of $\alpha(<0)$,
$x_{\rm B}$ decreases with increasing $K$, as expected.
The particle number density, $x_{\rm N}$, can be
readily computed from equation (\ref{eq:equiP_1}).

\subsection{Kinetic Luminosity}          
\label{sec:Lkin}

When the jet has a perpendicular half width $R_\perp$ at a
certain position,
$L_{\rm kin}$ and $\Nco$ are related by
\begin{equation}
   L_{\rm kin}
   =  \pi R_\perp{}^2 \beta c \cdot \Gamma \Nco \cdot (\Gamma-1)
      \left( \langle\gamma_-\rangle m_{\rm e}c^2 
            +\langle\gamma_+\rangle m_+ c^2 
      \right) ,
   \label{eq:Lkin}
\end{equation}
where $\langle\gamma_+\rangle$ refers to the 
averaged Lorentz factors of positively charged particles,
and $m_+$ designates the mass of the positive charge.

For a conical geometry, we can put $R_\perp=\rho\chi$.
Substituting this relation into equation~(\ref{eq:Lkin}),
and replacing $\Nco \rho^2$ with 
\begin{eqnarray}
  \Nco \rho^2 
    &=& N_1 r_1{}^2 = \frac{r_1}{r_\circ{}^2} x_{\rm N}
  \nonumber \\
    &=& \frac{\pi}{2} \frac{K}{\gamma_{\rm min}}
             \frac{r_1{}^2}{r_\circ{}^3} x_{\rm B}{}^2,
  \label{eq:NR2}
\end{eqnarray}
we obtain
\begin{eqnarray}
  L_{\rm kin}
  &=& C_{\rm kin} K 
      \frac{r_1{}^2}{r_\circ{}^3} \beta \Gamma (\Gamma-1)
      \left( \frac{1}{\nu_\circ}
             \frac{\chi\Omega_{r\nu}}{r_1 \sin\varphi}
      \right)^{2(5-2\alpha)/(7-2\alpha)}
  \nonumber \\
  && = \hspace{-1.5 truecm} \times
      \left[ \frac{\pi C(\alpha)}{\sin\varphi}
             \frac{K}{\gamma_{\rm min}}
             \frac{r_1}{r_\circ}
             \frac{-2\alpha}{\gamma_{\rm min}{}^{2\alpha}}
             \left(\frac{\delta}{1+z}\right)^{3/2-\alpha}
      \right]^{-4/(7-2\alpha)};
  \label{eq:Lkin_CPO}
\end{eqnarray}
we have
$C_{\rm kin}= \pi^2 \langle\gamma_-\rangle m_{\rm e}c^3/\gamma_{\rm min}$
for a pair plasma, while
$C_{\rm kin}= \pi^2 m_{\rm p}c^3/(2\gamma_{\rm min})$
for a normal plasma, 
where $m_{\rm p}$ refers to the rest mass of a proton.
It should be noted that $\gamma_{\rm min}$ for a pair plasma
takes a different value from that for a normal plasma.

\subsection{Jet Opening Angle}      
\label{sec:opening}

Let us now consider the half opening angle of the jet, 
and further reduce the expression of $\L_{\rm kin}$.
As noted in \S~\ref{sec:Lkin}, 
we obtain
\begin{equation}
  \chi = \frac{R_\perp}{\rho} 
       = \frac{0.5\theta_{\rm d,core}(\nu)}{\rho}
         \frac{D_{\rm L}}{(1+z)^2}
  \label{eq:chi}
\end{equation}
for a conical geometry,
where $\theta_{\rm d,core}(\nu)$ is the angular diameter of the VLBI core 
at a frequency $\nu$ in the perpendicular direction to the 
jet-propagation direction on a projected plane (i.e., a VLBI map).
The luminosity distance is given by
\begin{equation}
  D_{\rm L}= 4.61 \times 10^9 h^{-1} r_1
             \frac{q_0 z +(q_0-1)(-1+\sqrt{2q_0 z+1})}{q_0{}^2}.
  \label{eq:Da}
\end{equation}
On the other hand, equation (\ref{eq:Rcore}) gives
\begin{equation}
  \rho 
    = r_{\rm core} r_1
    = \frac{\Omega_{r \nu}}{\sin\varphi} \nu^{-1/k_{\rm r}}
  \label{eq:Rcore_2}
\end{equation}
Substituting equations~(\ref{eq:Rcore_2}), (\ref{eq:Da}) into 
(\ref{eq:chi}), we obtain
\begin{eqnarray}
  \frac{1}{\nu_0} \frac{\chi\Omega_{r\nu}}{r_1 \sin\varphi}
  &=& 11.1 h^{-1} \frac{q_0 z +(q_0-1)(-1+\sqrt{2q_0 z+1})}
                     {q_0{}^2 (1+z)^2}
  \nonumber\\
  & & \times
    \left( \frac{\theta_{\rm d,core}}{\rm mas} \right)
    \frac{\nu^{1/k_r}}{\nu_0}.
  \label{eq:chi_CPO}
\end{eqnarray}
As a result, the factor containing $\chi$ and $\Omega_{r\nu}$ 
in equation (\ref{eq:Lkin_CPO}) can be simply expressed 
in terms of $\theta_{\rm d,core}(\nu)$.
That is, to evaluate $L_{\rm kin}$,
we do not have to stop on the way at the computation of 
$\Omega_{r\nu}$,
which requires more than two VLBI frequencies.
We only need the VLBI-core size at a single frequency,
$\theta_{\rm d,core}(\nu)$,
in addition to $\Gamma$, $\varphi$, $\alpha$, $\gamma_{\rm min}$,
$z$, $K$, and $C_{\rm kin}$.
The information of the composition is included in $C_{\rm kin}$.
If we have to know $\chi$,
we additionally need $\Omega_{r\nu}$. 

\subsection{Electron density deduced from kinetic luminosity}
\label{sec:Ne_kin}

Once $L_{\rm kin}$ is known by the method described above
or by some other methods,
we can compute the electron number density as a function of the
composition.
Solving equation (\ref{eq:Lkin}) for $\Nco$,
and utilizing $R_\perp = (\theta_{\rm d}/2)D_{\rm L}/(1+z)^2$,
we obtain for a normal plasma
(i.e., for $\langle\gamma_+\rangle=1$ and $m_+=m_{\rm p}$)
\begin{eqnarray}
  \Nco({\rm nml})
  &=& 5.93 \times 10^{-2} h^2 \frac{L_{46}}{\Gamma(\Gamma-1)}
      \left( \frac{\theta_{\rm d}}{\rm mas} \right)^{-2}
  \nonumber \\
  & & \times \left[ \frac{q_0{}^2 (1+z)^2}
                     {zq_0+(q_0-1)(-1+\sqrt{2q_0z+1})}
         \right]^2
  \nonumber \\
  & & \times
         \frac{1}{1+\langle\gamma_-\rangle m_{\rm e}/m_{\rm p}}
         \, \mbox{cm}^{-3},
  \label{eq:Ne_nml}
\end{eqnarray}
where $L_{46}= L_{\rm kin}/(10^{46} \mbox{ergs s}^{-1})$.
If $\Nco({\rm nml})$ becomes less than the electron density deduced 
independently from the theory of synchrotron self-absorption (SSA),
the possibility of a normal plasma dominance can be ruled out.

It is worth comparing $\Nco({\rm nml})$ with $\Nco({\rm pair})$,
the particle density in a pure pair plasma.
If we compute $L_{\rm kin}$ by using equation~(\ref{eq:Lkin_CPO})
and (\ref{eq:chi_CPO}), we obtain the density ratio
\begin{eqnarray}
  \frac{\Nco({\rm pair})}{\Nco({\rm nml})}
  &=& \frac{L_{\rm kin}({\rm pair})}{L_{\rm kin}({\rm nml})}
      \frac{m_{\rm p}c^2 
            +\langle\gamma_-({\rm nml })\rangle m_{\rm e}c^2}
           {2\langle\gamma_-({\rm pair})\rangle m_{\rm e}c^2}
  \nonumber \\
  &=& \frac{\gamma_{\rm min}({\rm nml })}
           {\gamma_{\rm min}({\rm pair})},
  \label{eq:DensRat}
\end{eqnarray}
where $\gamma_{\rm min}({\rm pair})$ and $\gamma_{\rm min}({\rm nml})$
refer to $\gamma_{\rm min}$ when a pair and a normal plasma dominates,
respectively,
$\langle \gamma_-({\rm pair}) \rangle$ and 
$\langle \gamma_-({\rm nml})  \rangle$
refer to the averaged Lorentz factor of electrons in a
pair and a normal plasma, respectively.
Therefore, if 
$\gamma_{\rm min}({\rm pair})$ is comparable with 
$\gamma_{\rm min}({\rm nml}) \sim 100$,
there is little difference between 
$\Nco({\rm pair})$ and $\Nco({\rm nml})$.

In short, if $\Nco({\rm nml})$ becomes much less than
the $\Nco$ deduced independently by another method,
we can rule out the possibility of a normal-plasma dominance 
and that of a pair-plasma dominance with $\gamma_{\rm min} \sim 100$
or greater. 
In the next section, we investigate how to deduce 
$\Nco$ independently by the theory of SSA.

\section{Synchrotron Self-Absorption Constraints}      
\label{sec:SSA}

\subsection{Magnetic Field Strength}      
\label{sec:mag}

In this paper, we model a jet component
as a homogeneous sphere of angular diameter $\theta_{\rm d}$, 
containing a tangled magnetic field $B$ [G]
and relativistic electrons which give a synchrotron spectrum with
optically thin index $\alpha$ and maximum flux density $S_{\rm m}$
at the peak frequency $\nu_{\rm m}$.
We can then compute the magnetic field density as follows 
(Appendix):
\begin{equation}
  B = 10^{-5} b(\alpha) 
      \left( \frac{\nu_{\rm m}}{\rm GHz} \right)^5 
      \left( \frac{\theta_{\rm d}}{\rm mas} \right)^4
      \left( \frac{     S_{\rm m}}{\rm Jy}  \right)^{-2}
      \frac{\delta}{1+z},
  \label{eq:lineA}
\end{equation}
The coefficient $b(\alpha)$ is computed for a uniform sphere of
plasma with angular diameter $\theta_{\rm d}$;
their values are given in table~1. 

\begin{table*}
 \centering
 \begin{minipage}{140mm}
  \caption{\hspace{4pt}Table of constants}
  \begin{tabular}{@{}cccccccccc@{}}
  \hline
  \hline
  $\alpha$	& $0$
		& $-$0.25	& $-$0.50	
		& $-$0.75	& $-$1.00
		& $-$1.25	& $-$1.50	
		& $-$1.75	& $-$2.00	\\ 
  \hline
  $a$		& 0.2833 
		& 0.149		& 0.103
		& 0.0831	& 0.0740
		& 0.0711	& 0.0725
		& 0.0776	& 0.0865	\\
  $C$		& 1.191
		& 1.23		& 1.39
		& 1.67		& 2.09
		& 2.72		& 3.67
		& 5.09		& 7.23		\\
  $\tau_{\rm m}(0)$
		& 0
		& 0.187		& 0.354
		& 0.503		& 0.639
		& 0.762		& 0.876
		& 0.981		& 1.079		\\
  $\langle \tau_{\rm m} \rangle$
		& 0
		& 0.165		& 0.310
		& 0.440		& 0.557
		& 0.663		& 0.760
		& 0.848		& 0.930		\\
  $b$		& 0
		& 1.63		& 2.33	
		& 2.30		& 2.05
		& 1.75		& 1.48
		& 1.25		& 1.05		\\
  \hline
  \hline
\end{tabular}
\end{minipage}
\end{table*}



\subsection{Proper Electron Density}       
\label{Ne_SSA}

Substituting equation~(\ref{eq:abs-coeff}) into (\ref{eq:o-depth-2}),
and assuming $\gamma_{\rm min} \ll \gamma_{\rm max}$, we obtain
\begin{eqnarray}
  \Nco B^{-\alpha +1.5} 
  &=& \frac{m_{\rm e}c}{e^2}
      \left( \frac{e}{2\pi m_{\rm e}c} \right)^{-1.5+\alpha}
      \frac{\tau_\nu(\alpha)}{C(\alpha)}
      \frac{\gamma_{\rm min}{}^{2\alpha}}{-2\alpha}
  \nonumber \\
  &&  \hspace{-1.5 truecm} \times
      \frac{\sin \varphi}{\theta_{\rm d}}
             \frac{(1+z)^2}{D_{\rm L}}
      \left( \frac{1+z}{\delta} \right)^{-\alpha+1.5}
      \nu^{-\alpha+2.5}.
  \label{eq:lineB-1}
\end{eqnarray}
Evaluating $\nu$ at $\nu_{\rm m}$,
and combining with equation (\ref{eq:lineA}),
we obtain (Marscher 1983)
\begin{eqnarray}
   \Nco({\rm SSA})
   &=&  e(\alpha) 
      \frac{\gamma_{\rm min}{}^{2\alpha}}{-2\alpha}
      \frac{h (1+z)^2 q_0{}^2 \sin\varphi}
           {zq_0 +(q_0-1)(-1+\sqrt{2q_0 z+1})}
   \nonumber \\
   & & \hspace*{-2.0 truecm} \times
      \left( \frac{\theta_{\rm d}}{\rm mas} \right)^{4\alpha-7}
      \left( \frac{\nu_{\rm m}}{\rm GHz}    \right)^{4\alpha-5} 
      \left( \frac{S_{\rm m}}{\rm Jy}       \right)^{-2\alpha+3}
      \left( \frac{\delta}{1+z} \right)^{2\alpha-3},
   \label{eq:Nmin}
\end{eqnarray}
where 
\begin{equation}
  e(\alpha) \equiv
    1.71 \times 10^{-9}
    \times [ 2.79 \times 10^{-8} b(\alpha) ]^{\alpha-1.5}
    \times \frac{\tau_{\nu}}{C(\alpha)}
  \label{eq:def_e}
\end{equation}

It looks like from equation~(\ref{eq:Nmin}) that  
$\Nco({\rm SSA})$ 
depends on $\theta_{\rm d}$ and $\nu_{\rm m}$ very strongly.
In the case of $\alpha=-0.75$, for instance, we obtain 
$\Nco({\rm SSA}) \propto \theta_{\rm d}{}^{-10} \nu_{\rm m}{}^{-8}$.
Nevertheless, as a radio-emitting components evolves along the jet,
its $\theta_{\rm d}$ increases due to expansion
while its $\nu_{\rm m}$ decreases due to (synchrotron+adiabatic) cooling.
As a result, these two effects partially cancle each other and
reduce the variations of $\Nco({\rm SSA})$ along the jet.

\subsection{Summary of the Method}

Let us summarize the main points that have been made in \S\S~2 and 3. \\
(1) \ 
Observing the core size, $\theta_{\rm d,core}(\nu)$, of a conical jet
at a single frequency, $\nu$,
we obtain
\begin{equation}
  \frac{1}{\nu_\circ}
  \frac{\chi \Omega_{r\nu}}{r_1 \sin\varphi}
  \propto \theta_{\rm d,core} \cdot \nu^{1/k_r}
  \label{eq:sum_1}
\end{equation}
by equation~(\ref{eq:chi_CPO}).\\
(2) \ 
Substituting equation~(\ref{eq:chi_CPO}) into (\ref{eq:Lkin_CPO}),
we obtain the kinetic luminosity, $L_{\rm kin}$,
in terms of $\theta_{\rm d,core} \cdot \nu^{1/k_r}$, $\alpha$,
$\Gamma$, $\varphi$, $K$ and $\gamma_{\rm min}$ for the core.
Note that $\gamma_{\rm min}$ takes different values 
between a normal plasma and a pair plasma.\\
(3) \ 
Assuming a normal-plasma dominance,
we obtain the proper electron density, $\Nco({\rm nml})$
from $L_{\rm kin}$ by equation~(\ref{eq:Ne_nml}).\\
(4) \ 
We can independently deduce $\Nco$ from the theory of SSA 
as $\Nco({\rm SSA})$ by equation~(\ref{eq:Nmin}).\\
(5) \ 
If $\Nco({\rm nml}) \ll \Nco({\rm SSA})$ holds,
we can rule out the possibility of a normal-plasma dominance and 
that of a pair-plasma dominance with $\gamma_{\rm min} \sim 100$ or greater.
That is, $\Nco({\rm nml}) \ll \Nco({\rm SSA})$ indicates the
dominance of a pair plasma with $\gamma_{\rm min} \ll 100$.


\section{Application to the 3C~345 Jet}    
\label{sec:3C345_appli}

Let us apply the method described in the previous sections
to the radio-emitting components in the 3C~345 jet
and investigate the composition.
This quasar ($z=0.595$; Hewitt \& Burbidge 1993)
is one of the best studied objects showing structural and spectral
variabilities on parsec scales around the compact unresolved core
(for a review, see e.g., Zensus 1997).
At this distance, 1 mas corresponds to $5.42 h^{-1}$ pc,
and 1 mas ${\rm yr}^{-1}$ to $\beta_{\rm app}= 30.3 h^{-1}$,
where $H_0 = 65 h {\rm km s}^{-1}{\rm Mpc}^{-1}$.

\subsection{Kinetic luminosity}
\label{sec:3C345_Lkin}

\begin{figure} 
\centerline{ \epsfxsize=9cm \epsfbox[200 2 500 350]{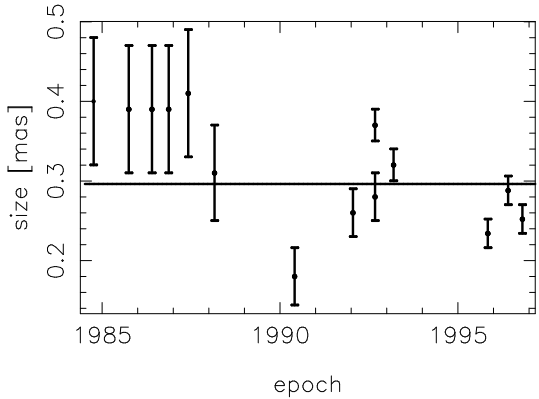} } 
\caption{\label{fig:core22} 
Angular core size of 3C~345 at 22.2 GHz.
The horizontal line represents the bet fit value.
        }
\end{figure}

To deduce $L_{\rm kin}$, we first consider 
$\theta_{\rm d, core} \nu / \nu_0$, 
where $k_r=1$ is adopted in equation~(\ref{eq:chi_CPO}).
From the reported core size at 22.2 GHz
    at 6 epochs by Zensus et al. (1995),
    at 5 epochs by Unwin  et al. (1997),
and at 3 epochs by Ros    et al. (2000),
we can deduce the averaged core size, $\theta_{\rm d,core}$,
as $0.296$ mas (fig.~\ref{fig:core22});
here, we evaluate the core size with $1.8\sqrt{ab}$ for
the latter three epochs, 
where $a$ and $b$ refer to the major and minor axes at the FWHM
of the elliptical Gaussian presented in Ros et al. (2000).
From $\theta_{\rm d,core}= 0.296$ mas, we obtain 
$\theta_{\rm d,core} \nu / \nu_0 = 6.18 \times 10^{-14}$
as the averaged value over the 14 epochs at 22 GHz.
In the same manner, we obtain the following averaged values
at different frequencies:
$\theta_{\rm d,core}= 0.258$ mas and 
$\theta_{\rm d,core} \nu / \nu_0 =  3.64 \times 10^{-14}$ 
over 3 epochs at 15 GHz (Ros et al. 2000);
$\theta_{\rm d,core}= 0.436$ mas and 
$\theta_{\rm d,core} \nu / \nu_0 =  4.38 \times 10^{-14}$ 
over 7 epochs at 10.7 GHz 
(1 epoch from Unwin et al. 1994;
 4 epochs from Zensus et al. 1995; 
 2 epochs from Gabuzda et al. 1999);
$\theta_{\rm d,core}= 0.343$ mas and 
$\theta_{\rm d,core} \nu / \nu_0 =  2.71 \times 10^{-14}$ 
over 7 epochs at 8.4 GHz 
(1 epoch from Unwin et al. 1994;
 3 epochs from Zensus et al. 1995; 
 3 epochs from Ros et al. 2000);
$\theta_{\rm d,core}= 0.423$ mas and 
$\theta_{\rm d,core} \nu / \nu_0 =  1.98 \times 10^{-14}$ 
over 11 epochs at 5.0 GHz 
(4 epochs from Brown et al. 1994;
 1 epoch from Unwin et al. 1994;
 3 epochs from Zensus et al. 1995;
 3 epochs from Ros et al. 2000).
Taking the weighted average of 
$\theta_{\rm d,core} \nu / \nu_0$ 
over the 42 epochs at the different 5 frequencies, we obtain
\begin{equation}
  \frac{\theta_{\rm d,core} \nu}{\nu_0}= 4.01 \times 10^{-14}.
  \label{eq:core_1}
\end{equation}

We next consider the spectral index
that reflects the energy distribution of the power-law electrons
in the core.  
From infrared to optical observations 
(J,H,K,L bands from Sitko et al. 1982;
 J,H,K   bands from Neugebauer et al. 1982;
 K       band  from Impey et al. 1982;
 IRAS 12, 25, 60, 100 microns from Moshir et al. 1990),
we obtain 
\begin{equation}
  \alpha= -1.15
  \label{eq:core_2}
\end{equation}
for the core.
This value is consistent with that adopted by Zensus et al. (1995).

Using equations~(\ref{eq:core_1}) and (\ref{eq:core_2}), we obtain
from equation~(\ref{eq:chi_CPO})
\begin{equation}
  \left( \frac{1}{\nu_0} 
         \frac{\chi \Omega_{r\nu}}{r_1 \sin\varphi}
  \right)^{2(5-2\alpha)/(7-2\alpha)}
  = 4.97 \times 10^{-21} h^{-1.57}.
  \label{eq:core_3}
\end{equation}
Moreover, $\alpha=-1.15$ gives
\begin{eqnarray} 
  \, & &
  \left[ \pi C(\alpha) \frac{-2\alpha K}{\gamma_{\rm min}{}^{2\alpha+1}}
         \frac{r_1}{r_0}
  \right]^{-4/(7-2\alpha)}
  \nonumber \\
  &=& 1.31 \times 10^{-14} K^{-0.43} \gamma_{\rm min}{}^{-0.56}.
  \label{eq:core_4}
\end{eqnarray}
We thus obtain 
\begin{eqnarray}
  L_{\rm kin}
  &=& 4.68 \times 10^{45} h^{-1.57} 
      \left(\frac{\gamma_{\rm min}}{100}\right)^{-1.56} K^{0.57}
  \nonumber\\
  & & \times \Gamma(\Gamma-1)
             \left[ \frac{1}{\sin\varphi}
                    \left(\frac{\delta}{1+z}\right)^{2.65}
             \right]^{-0.43}
      \mbox{ergs s}^{-1}
  \label{eq:Lkin_3C345_0}
\end{eqnarray}
for a normal plasma, that is, 
$C_{\rm kin}= \pi^2 m_{\rm p}c^3/(2\gamma_{\rm min})$
in equation~(\ref{eq:Lkin_CPO}).
             
\subsection{Kinematics near the core}
\label{sec:3C345_kinematics}

To constrain $L_{\rm kin}$ further,
we must consider the kinematics of the core
and infer $\Gamma$ and $\varphi$. 
If the apparent component velocity, $\beta_{\rm app}$,
reflects the fluid velocity,
the lower bound of $\Gamma$ can be obtained by 
$\Gamma_{\rm min} \equiv \sqrt{\beta_{\rm app}{}^2+1}$.

In the case of the 3C~345 jet,
components are found to be \lq accelerated' as they propagate
away from the core.
For example, for component C4,
$\beta_{\rm app}h<8$ held when the distance from the core was 
less than 1 mas (Zensus et al. 1995);
however, it looked to be '\lq accelerated' from this distance
(or epoch 1986.0)
and attained $\beta_{\rm app}h \sim 45$
at epoch 1992.5 (Lobanov \& Zensus 1994).
It is, however, unlikely that the fluid bulk Lorentz factor increases
so much during the propagation.
If $\beta_{\rm app}$ reflected the fluid bulk motion of C4,
the bulk Lorentz factor, $\Gamma$, had to be at least $45$ 
until 1992.5.
Then $\beta_{\rm app} \sim 6$ during 1981-1985 indicates
that the viewing angle should be less than $0.05^\circ$,
which is probably much less than the opening angle. 
Analogous \lq acceleration' was reported also for C3 (Zensus et al. 1995) 
and C7 (Unwin et al. 1997). 

On these grounds, 
we consider the apparent \lq acceleration' of the components 
in their later stage of propagation on VLBI scales
are due to geometrical effects such as \lq scissors effects'
(Hardee \& Norman 1989; Fraix-Burnet 1990),
which might be created at the intersection of a pair of shock waves.
Such shock waves may be produced, for example, 
by the interaction between relativistic pair-plasma beam and 
the ambient, non-relativistic normal-plasma wind 
(Pelletier \& Roland 1989; Pelletier \& Sol 1992).  
In their two-fluid model,
the pair beam could be destructed 
at a certain distance from the core on VLBI scales
due to the generation of Langmuir waves.

Then, how should we constrain $\Gamma$ and $\varphi$ for the 3C~345 jet?
In this paper, we assume that the radio-emitting components 
before their rapid acceleration represent 
(or mimic) the fluid bulk motion.
Under this assumption,
Steffen et al. (1995) fitted the motions of C4 
(during 1980-1986) and C5 (during 1983-1989)
when the components are within 2 mas from the core by a helical model.
They found $\Gamma= 5.8$ for C4 and $4.6$ for C5
with $\varphi= 6.8^\circ$, assuming $h=1.54$ (i.e., $H_0= 100$km/s/Mpc).
These Lorentz factors will be $\sim 9$ and $\sim 7$ for C4 and C5, 
respectively,
if $h=1$ (i.e., $H_0=65$km/s/Mpc).
Subsequently, Qian et al. (1996) investigated the intrinsic evolution
of C4 under the kinematics of $\Gamma=5.6$ and 
$\varphi=2^\circ$ -- $8^\circ$, assuming h= 1.54;
the range of viewing angles is in good agreement with 
$\varphi=2^\circ$ (for C4) -- $4^\circ$ (for C2) at 1982.0
obtained by Zensus et al. (1995), 
who used a larger value of $\Gamma=10$ under $h=1.54$,
which corresponds to $\Gamma \sim 15$ for $h=1$.

In the present paper, we assume that the helical model 
describes the fluid bulk motion when the components are within
2 mas from the core
and adopt $\Gamma=15$ as the typical value.
If $\Gamma$ is less than $15$, 
the possibility of a normal-plasma dominance further decreases.
Close to the core, we adopt $\varphi=2^\circ$,
because the viewing angle is suggested to decrease with decreasing
distance from the core (Zensus et al. 1995; Unwin et al. 1997).
If $\varphi$ is less than $2^\circ$ in the core,
$L_{\rm kin}$, and hence $\Nco({\rm nml})$ further decreases;
that is, the normal-plasma dominance can be further ruled out.
For $\Gamma=15$ and $\varphi=2^\circ$ 
(or equivalently $\varphi_{\rm A}= 1.3^\circ$), 
we obtain $\delta= 27.0$ for the core and hence
\begin{equation}
  L_{\rm kin}
    = 9.2 \times 10^{45} h^{-1.57} K^{0.57}
      \left( \frac{\gamma_{\rm min}}{100} \right)^{-1.56}
      \mbox{ergs s}^{-1},
  \label{eq:Lkin_3C345_1}
\end{equation}
assuming a normal plasma.
Substituting equation~(\ref{eq:Lkin_3C345_1}) into (\ref{eq:Ne_nml}),
we obtain $\Nco({\rm nml})$ for each component having
angular diameter $\theta_{\rm d}$.

\subsection{Composition of individual components}
\label{sec:3C345_individual}

We can now investigate the composition of the radio-emitting 
components in the 3C~345 jet, using their spectral information. 
We assume that the jet is neither accelerated nor decelerated and 
apply a constant Lorentz factor of $\Gamma=15$ for all the components, 
from the reasons described just above.
To compute $\Nco({\rm SSA})$, we further need $\varphi$
for each component at each epoch.
(Note that $\varphi=2^\circ$ is assumed for the core, 
not for the components.)
For this purpose, we assume that $\beta_{\rm app}$ reflects
the fluid motion if $\beta_{\rm app}<\sqrt{\Gamma^2-1}$ and 
that $\varphi$ is fixed after $\beta_{\rm app}$
exceeds $\sqrt{\Gamma^2-1}$.
That is, we compute $\varphi$ (or equivalently, $\varphi_{\rm A}$) from
\begin{eqnarray}
  \tan\varphi_{\rm A}
    &=& \frac{2\beta_{\rm app}}{\beta_{\rm app}{}^2+\delta^2-1}
        \qquad \mbox{if} \quad \beta_{\rm app}<\sqrt{\Gamma^2-1}
    \nonumber \\
    &=& \frac{1}{\sqrt{\Gamma^2-1}}
        \qquad \mbox{if} \quad \beta_{\rm app}>\sqrt{\Gamma^2-1}
    \label{eq:view_ang}
\end{eqnarray}
The constancy of $\varphi$ (or $\varphi_{\rm A}$)
when $\beta_{\rm app}>\sqrt{\Gamma^2-1}=15.0$ could be justified 
by the helical model of Steffen et al. (1995),
who revealed that the viewing angle does not vary so much 
after the component propagates a certain distance 
(about 1 mas in the cases of components C4 and C5) from the core.

The synchrotron self-absorption spectra of individual components are
reported in several papers.
They were first reported in Unwin et al. (1994),
who presented ($\nu_{\rm m}$,$S_{\rm m}$,$\alpha$) and $\theta_{\rm d}$
(or $\xi$ in their notation) of components C4 and C5 at epoch 1990.7.
Subsequently, Zensus et al. (1995) gave 
($\nu_{\rm m}$,$S_{\rm m}$,$\alpha$), $\theta_{\rm d}$, 
and $\beta_{\rm app}h$ 
for C2, C3, and C4 at 1982.0.
However, the errors are given only for $\theta_{\rm d}$; therefore,
we evaluate the errors of the output parameters,
$\delta$, $\varphi$, $B$, $\Nco({\rm nml})$, 
and $\Nco({\rm nml})/\Nco({\rm SSA})$,
using the errors in $\theta_{\rm d}$ alone for C2, C3, and C4 at 1982.0.
Later, Unwin et al. (1997) presented
($\nu_{\rm m}$,$S_{\rm m}$,$\alpha$), $\theta_{\rm d}$, 
and $\beta_{\rm app}h$ 
of C5 (at 1990.55) and C7 (at 1992.05, 1992.67, 1993.19, and 1993.55).
More recently, Lobanov and Zensus (1999) presented a comprehensive data
of ($\nu_{\rm m}$,$S_{\rm m}$,$\alpha$) of C3, C4, and C5 at various
epochs.
We utilize $\theta_{\rm d}= 0.114 + 0.0658({\rm epoch}-1979.50)$ for C4
(Qian et al. 1996),
$\theta_{\rm d}= 0.114 + 0.0645({\rm epoch}-1980.00)$ for C5,
which is obtained from the data given in Zensus et al. (1995) and 
Unwin et al. (1997),
and $\theta_{\rm d}= 0.09 + 0.45(\rho\sin\varphi/{\rm mas})$ for C3 
(Biretta et al. 1986),
where $\rho\sin\varphi$ refers to the projected distance from the core.

The results for individual components at various epochs are given
in table~2 for $K=1$.
It follows that the ratio, $\Nco({\rm nml}) / \Nco{}({\rm SSA})$ 
becomes less than $1$ for acceptable strength of the 
magnetic field, that is, $1{\rm mG}<B<100{\rm mG}$, say.
For C4, we cannot rule out the possibility of a normal-plasma dominance
at epochs 1985.8 and 1988.2.
Nevertheless, at these two epochs, 
the input parameters 
($\nu_{\rm m}$, $S_{\rm m}$, $\alpha$, $\theta_{\rm d}$)
give too large $B$ $(>380 \mbox{mG})$ for a moderate $\delta (<30)$.
For C5 at epoch 1984.2, 
the input parameters give too small $B$ ($= 9 \mu \mbox{G}$) for 
a moderate $\delta(=29.1)$.
On these grounds, we discard these three cases.

\begin{figure} 
\centerline{ \epsfxsize=9cm \epsfbox[200 2 500 350]{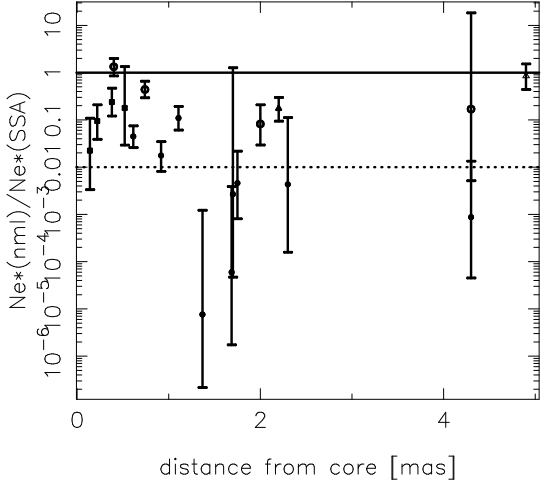} } 
\caption{\label{fig:ratio_1} 
The ratio $\Nco({\rm nml}) / \Nco{}({\rm SSA}) $ 
as a function of the distance from the core.
Above the solid, horizontal line, the dominance of a normal plasma
is allowed.
$\Gamma=15$ is assumed throughout the jet.
        }
\end{figure}

We present the ratio, $\Nco({\rm nml}) / \Nco{}({\rm SSA})$ 
as a function of $\rho\sin\varphi$ in figure~\ref{fig:ratio_1}.
The solid line represents $\Nco=\Nco({\rm nml})$.
It should be noted that $\Nco({\rm nml})$ is computed under
$\langle\gamma_-\rangle=0$ in equation~(\ref{eq:Ne_nml}).
Thus, the solid line gives the conservative upper limit of 
$\Nco({\rm nml}) / \Nco{}({\rm SSA})$. 
In another word, if $\Nco({\rm nml}) / \Nco{}({\rm SSA})$ slightly
exceeds unity ($1.5$, say), 
a normal-plasma dominance can be, in general, ruled out.
There is also depicted a dotted line representing $\Nco=\Nco({\rm pair})$.
If the ratio appears under the dotted line,
it indicates that $L_{\rm kin}$ is underestimated or that
the input parameters give too small $B$
and hence too large $\Nco{}({\rm SSA})$.

It follows from the figure that neither a normal-plasma dominance 
nor the pair-plasma dominance with $\gamma_{\rm min} > 100$ 
are allowed within one-$\sigma$ errors
for the 14 epochs out of the 19 epochs at which input parameters
give ordinary $B$.
The ratio shows
no dependence on $\rho$ under the assumption of constant $\Gamma$.

It is noteworthy that 
$\Nco({\rm nml})/\Nco({\rm SSA}) 
 \propto (\gamma_{\rm min}/100)^{-2\alpha-1.56}$
for the spectral index of $-1.15$ for the core.
Therefore, for a typical spectral index $\alpha \sim -0.75$
for the jet components, 
the dependence on $\gamma_{\rm min}$ virtually vanishes.
That is, the conclusion does not depend on the assumed value of 
$\gamma_{\rm min}\sim 100$ for a normal plasma.

\begin{figure} 
\centerline{ \epsfxsize=9cm \epsfbox[200 2 500 350]{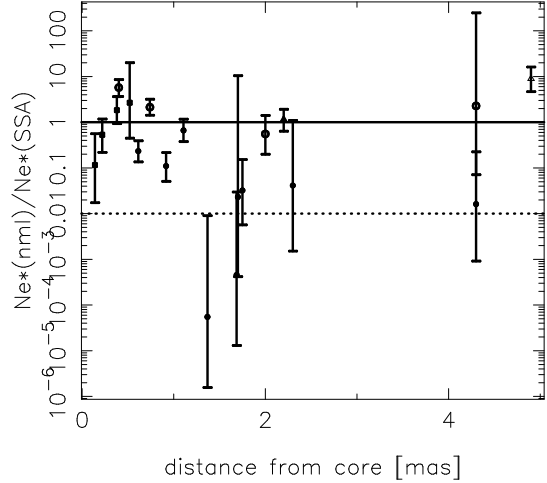} } 
\caption{\label{fig:ratio_2} 
The same figure as figure~3,
but the different Lorentz factor $\Gamma=20$ is adopted.
        }
\end{figure}

If we assume instead $\Gamma=20$ (rather than $\Gamma=15$)
and $\varphi=2^\circ$,
we obtain the following kinetic luminosity
\begin{equation}
  L_{\rm kin}= 1.3 \times 10^{46} h^{-1.57}
               \left( \frac{\gamma_{\rm min}}{100} \right)^{-1.56}
               K^{0.56}
               \mbox{ergs s}^{-1}.
  \label{eq:Lkin_3C279_4}
\end{equation}
Evaluating the viewing angles of individual components by
equation~(\ref{eq:view_ang}) with $\Gamma=20$,
and excluding the three epoch data 
(C4 at 1985.8 and 1988.2, and C5 at 1984.2) that give extraordinary $B$,
we can compute $\Nco({\rm nml}) / \Nco({\rm SSA})$ as
presented in figure~\ref{fig:ratio_2}. 
It follows from this figure that we cannot rule out the possibility
of a normal-plasma dominance for such large Lorentz factors.
The greater $\Gamma$ we assume,
the greater becomes $\Nco({\rm nml}) / \Nco({\rm SSA})$.

On these grounds, we can conclude that the jet is dominated
by a pair plasma with $\gamma_{\rm min} \ll 100$
at 14 epochs out of the 19 epochs at which input parameters 
give extraordinary $B$,
if we assume $\Gamma<15$ throughout the jet.


\section{Application to the 3C~279 Jet}    
\label{sec:3C279_appli}

Let us next consider the 3C~279 jet.
This quasar ($z=0.538$)
was the first source found to show superluminal motion 
(Cotton et al. 1979; Unwin et al. 1989).
At this distance, 1 mas ${\rm yr}^{-1}$ 
corresponds to $\beta_{\rm app}= 28.2 h^{-1}$.

\subsection{Kinetic luminosity}
\label{sec:3C279_Lkin}

In the same manner as in 3C~345 case,
we first consider $\theta_{\rm d, core} \nu / \nu_0$.
Since the turnover frequency of the core is reported to be about
$13$ GHz (Unwin et al. 1989),
we measure $\theta_{\rm d,core}$ at $\nu \leq 10.7$ GHz.
At 10.7 GHz,
$\theta_{\rm d,core}$ is reported to be
$0.79$ mas from five-epoch observations by Unwin et al. (1989), 
and $0.61$ mas from three-epoch observations by Carrara et al. (1993).
With one-epoch VSOP (VLBI Space Observatory Programme) observation,
$\theta_{\rm d,core}= 1.8 \times 0.28$ mas and 
$\theta_{\rm d,core}= 1.8 \times 1.95$ mas were obtained
at 4.8 GHz and 1.6 GHz, respectively (Piner et al. 2000).
As the weighted mean over the 10 epochs, we obtain
\begin{equation}
  \frac{\theta_{\rm d,core} \nu}{\nu_0}= 6.6 \times 10^{-14},
  \label{eq:core_10}
\end{equation}
which leads to
\begin{eqnarray}
  \frac{1}{\nu_0} \frac{\chi \Omega_{r\nu}}{r_1 \sin\varphi}
  &=& 2.8 h^{-1} \left(\frac{\theta_{\rm d,core}}{\rm mas}\right)
                 \frac{\nu}{\nu_\circ}
  \nonumber \\
  &=& 3.7 \times 10^{-13}. 
  \label{eq:core_11}
\end{eqnarray}

We next consider the spectral index of the core in the optical thin
frequencies.
From mm to IRAS (90-150 GHz and 270-370 GHz) observations, 
Grandi et al. (1996) reported 
\begin{equation}
  \alpha= -1.2.
  \label{eq:core_12}
\end{equation}
Adopting this value as the spectral index that
reflects the energy distribution of the power-law electrons in the core,
we obtain
\begin{equation}
  \left( \frac{1}{\nu_0} 
         \frac{\chi \Omega_{r\nu}}{r_1 \sin\varphi}
  \right)^{2(5-2\alpha)/(7-2\alpha)}
  = 2.7 \times 10^{-20} h^{-1.57}.
  \label{eq:core_13}
\end{equation}
and 
\begin{eqnarray} 
  \, & &
  \left[ \pi C(\alpha) \frac{-2\alpha K}{\gamma_{\rm min}{}^{2\alpha+1}}
         \frac{r_1}{r_0}
  \right]^{-4/(7-2\alpha)}
  \nonumber\\
  &=& 1.81 \times 10^{-14} K^{-0.43} \gamma_{\rm min}{}^{-0.56}.
  \label{eq:core_14}
\end{eqnarray}
If we assume a normal plasma dominance, we obtain
\begin{eqnarray}
  L_{\rm kin}
  &=& 3.0 \times 10^{46} h^{-1.57} 
      \left(\frac{\gamma_{\rm min}}{100}\right)^{-1.56} K^{0.57}
  \nonumber\\
  &\times& \Gamma(\Gamma-1)
             \left[ \frac{1}{\sin\varphi}
                    \left(\frac{\delta}{1+z}\right)^{2.7}
             \right]^{-0.43}
      \mbox{ergs s}^{-1}.
  \label{eq:Lkin_3C279_0}
\end{eqnarray}

\subsection{Kinematics near the core}
\label{sec:3C279_kinematics}

We now consider $\Gamma$ and $\varphi$ near to the core 
to constrain $L_{\rm kin}$ further.
The apparent superluminal velocities of the components in the 3C~279 jet
do not show the evidence of acceleration 
and are roughly constant during the propagation.
For example, Carrara et al. (1993) presented 
$\beta_{\rm app}= 4.51 h^{-1}$ for C3 during 1981 and 1990 over 15 epochs
at 5, 11, and 22 GHz
and 
$\beta_{\rm app}= 4.23 h^{-1}$ for C4 during 1985 and 1990 over 15 epochs
at 11 and 22 GHz.
We thus assume that the apparent motion of the components 
represent the fluid bulk motion
and that both $\Gamma$ and $\varphi$ are kept constant
throughout this straight jet.
Since $\beta_{\rm app}$ does not differ very much between C3 and C4,
we adopt $\beta_{\rm app}= 4.51$ as the common value 
and apply the same $\Gamma$ and $\varphi$ for these two components. 
The errors incurred by the difference in $\beta_{\rm app}$ are small.

Let us first consider the lower bound of $\Gamma$.
From the observed $\beta_{\rm app}$, we obtain
$\Gamma > \sqrt{\beta_{\rm app}{}^2 +1}= 4.61$.
As case~1, we adopt $\Gamma=4.61$ and $\beta_{\rm app}=4.51$.
It follows that 
$\varphi_{\rm A}= 12.5^\circ$ (or $\varphi=8.1^\circ$) and 
$\delta= 4.6$.
It is worth comparing this $\delta$ with that measured in the 
quiescent state; 
Unwin derived $\delta= 3.6 {+4.0 \atop -1.6}$ so that
the calculated X-ray flux with this $\delta$ agrees with the observed value.
Moreover, from the RTXE observation in the 0.1-2.0 keV band
during 1996.06 and 1996.11,
Lawson and M${}^{\rm c}$Hardy (1998) 
obtained $2 \mu$Jy in the quiescent state 
during 1996.49--1996.54 from 16-epoch observations.
Applying this X-ray flux density to the spectral information 
($\nu_{\rm m}$, $S_{\rm m}$, $\alpha$) and $\theta_{\rm d}$
(table~3) for component C4 at epochs 1987.40 and 1989.26
and component C3 at 1984.05, 
we obtain $\delta > 1.9$, $>4.0$, and $>3.3$, respectively.
Therefore, we can regard that 
the value of $\delta=4.6$ are consistent with the X-ray observations.
In this case, equation~(\ref{eq:Lkin_3C279_0}) gives
\begin{equation}
  L_{\rm kin}
  = 6.3 \times 10^{46} K^{0.57}
    \left(\frac{\gamma_{\rm min}}{100}\right)^{-1.6}
    h^{-1.6}
    \, \mbox{ergs s}^{-1} \, \mbox{(case~1)}.
  \label{eq:Lkin_3C279_1}
\end{equation}
This case will be examined in the next subsection.

Let us next consider the upper bound of $\Gamma$,
which can be constrained by the upper bound of $\delta$
for a given $\beta_{\rm app}= 4.51$.
In general, the upper bound of $\delta$ is difficult to be deduced.
In the case of 3C~279, however, 
it is reasonable to suppose that a substantial fraction of 
the X-ray flux in the flaring state is produced via synchrotron
self Compton (SSC) process.
Therefore, the Doppler factors
$\sim 3.9$ (Mattox et al. 1993) and
$\sim 5  $ (Henri et al.  1993) derived for the 1991 flare,
and $\sim 6.3$ (Wehrle et al. 1998) for the X-ray flare in 1996,
give good estimates.
We thus consider that $\delta$ does not greatly exceed 10 for
the 3C~279 jet.
Imposing $\delta<10$, we obtain $\Gamma < 6.06$
from $\beta_{\rm app}=4.61$.
Combining with the lower bound, 
we can constrain the Lorentz factor in the range 
$4.6 < \Gamma < 6.1$.

As case~2, we adopt a large Lorentz factor of $\Gamma=6.0$.
In this case, $\beta_{\rm app}= 4.51$ gives
$\delta=9.8$ and $\varphi_{\rm A}= 4.4^\circ$ 
(or equivalently, $\varphi=6.8^\circ$).
There is, in fact, another branch of solution 
that gives smaller Doppler factor of $2.2$; 
however, this solution gives too large 
$\Nco({\rm SSA})$'s ($\sim 10^5 \mbox{cm}^{-3}$).
We thus consider only the solution giving $\delta=9.8$ as case~2,
which results in 
\begin{equation}
  L_{\rm kin}
  = 4.3 \times 10^{46} K^{0.57}
    \left(\frac{\gamma_{\rm min}}{100}\right)^{-1.6}
    h^{-1.6}
    \, \mbox{ergs s}^{-1} \, \mbox{(case~2)}.
  \label{eq:Lkin_3C279_2}
\end{equation}
We will consider this case, together with case~1, in the next subsection.

\subsection{Composition of individual components}
\label{sec:3C279_individual}

For the jet component C3, 
Unwin et al. (1989) presented $\nu_{\rm m}= 6.8$ GHz,
$S_{\rm m}= 9.4$ Jy, $\alpha= -1.0$, and $\theta_{\rm d}= 0.95$ mas
at epoch 1983.1 (table~3).
It was also reported in their paper that the flux densities were
$5.13 \pm 0.14$ Jy at 5.0 GHz  (at epoch 1984.25),
$4.49 \pm 0.17$ Jy at 10.7 GHz (1984.10), and 
$2.58 \pm 0.18$ Jy at 22.2 GHz (1984.09) for C3.
We modelfit the three-frequency data by the function
\begin{equation}
  S_\nu = A_1 \left[ 1-\exp(-A_2 \nu^{\alpha-2.5}) \right].
  \label{eq:model_SSA}
\end{equation}
Assuming $\alpha= -1.0$, we obtain
$\nu_{\rm m}= 6.6$ GHz and $S_{\rm m}= 5.9$ Jy as the best fit
around epoch 1984.10.

For C4, Carrara et al. (1993) presented 
$\nu_{\rm m} \sim 11 $ GHz,
$S_{\rm m} \sim 4.3$ Jy, $\alpha= -0.9$, and $\theta_{\rm d} \sim 0.6$ mas
from their 1989-1990 maps at 5, 11, and 22 GHz.
At epoch 1987.4,
we use the flux density of $1.43 \pm 0.17$ Jy at 22 GHz 
(Carrara et al. 1993) 
and that of $3.95 \pm 0.20$ Jy at 5 GHz (Gabuzda et al. 1999).
We extrapolate the flux density at 11 GHz from those at 
1988.17, 1989.26, and 1990.17 presented in Carrara et al. (1993)
to obtain $3.60 \pm 0.20$ Jy.
Assuming $\alpha=-0.9$, we can model fit the three-frequency
data to obtain
$\nu_{\rm m}=6.4$ GHz and $S_{\rm m}=4.4$ Jy.
At this epoch (1987.4), 
component C4 is located about 1 mas from the core;
therefore, the angular size vs. distance relation (Carrara et al. 1993) 
gives $\theta_{\rm d} \sim 0.6$ mas.
The input parameters for C3 and C4 at these 4 epochs are
summarized in table~3.

In case~1 (see \S~\ref{sec:3C279_kinematics}), 
we obtain $\Nco({\rm SSA})= 6.4 \times 10^4$ and $9.4 \times 10^3$, 
for C3 at 1983.1 and 1984.10,
and $1.2 \times 10^4$ and $9.7 \times 10^1$ 
for C4 at 1987.4 and 1989.26, respectively.
These large values of $\Nco({\rm SSA})$ result in such (unphysically)
small values of $\Nco({\rm nml})/\Nco({\rm SSA})$ as
$9.6 \times 10^{-5}$ for C3 at 1983.1,
$1.9 \times 10^{-5}$ for C3 at 1984.1, and
$1.3 \times 10^{-4}$ for C4 at 1987.4.
Since even a pair-plasma dominance should be ruled out when
$\Nco({\rm nml})/\Nco({\rm SSA}) \ll 100$ holds 
(see eq.~[\ref{eq:DensRat}]),
we consider that the value of $\delta(=4.6)$ is underestimated.
We thus exclude case~1 from consideration.

We next examine case~2, in which a larger Doppler factor ($=9.8$) 
is adopted.
In this case, we obtain reasonable values of $B$ and $\Nco({\rm SSA})$
(table~3).
It follows that the bestfit values of 
$\Nco({\rm nml})/\Nco({\rm SSA})$ becomes
$2.8 \times 10^{-3}$ and $5.7 \times 10^{-3}$, 
for C3 at 1983.1 and 1984.10,
and $3.2 \times 10^{-3}$ and $0.40$ 
for C4 at 1987.4 and 1989.26, respectively.
If $\delta$ exceeds $10$ (or equivalently, if $\Gamma$ exceeds $6$),
the density ratio for C4 at 1989.26 becomes close to unity;
therefore, we cannot rule out the possibility of a normal-plasma 
dominance for such large Doppler factors.

On these grounds, 
we can conclude that the jet components are dominated
by a pair plasma with $\gamma_{\rm min} \ll 100$,
provided that $\delta<10$ holds.
However, if $\delta$ is as small as $5$,
not only a normal-plasma dominance but also a pair-plasma dominance
is ruled out.
Thus, we consider that $\delta < 5$ is unphysical for the 3C 379 jet.

\section{Discussion}
\label{sec:discussion}

In summary, we derive the proper electron number density, 
$\Nco({\rm SSA})$, 
of a homogeneous radio-emitting component
of which spectral turnover is due to synchrotron self-absorption.
Comparing $\Nco({\rm SSA})$ with the density derived from the 
kinetic luminosity of the jet, we can investigate whether 
the possibility of a normal plasma ($e^-$-p) dominance
can be ruled out or not. 
Applying this method to the 3C~345 jet, 
and using the published spectral data of components 
C2, C3, C4, C5, and C7,
we find that they are dominated by a pair plasma with
$\gamma_{\rm min} \ll 100$ at 14 epochs out of the total 19 epochs,
if we assume $\Gamma<15$ throughout the jet.
We also investigate the 3C~279 jet
and find that components C3 and C4 are dominated
by a pair plasma with $\gamma_{\rm min} \ll 100$
at all the four epochs,
provided that $\delta < 10$. 

It is noteworthy that the kinetic luminosity computed from 
equations~(\ref{eq:Lkin_CPO}) and (\ref{eq:chi_CPO})
has, in fact, weak dependence on the composition. 
Substituting $C_{\rm kin}$ for a pair and a normal plasma, we obtain
\begin{eqnarray}
  \frac{L_{\rm kin}({\rm pair})}{L_{\rm kin}({\rm nml})}
  &=& 
  \frac{m_{\rm e} \langle\gamma_-({\rm pair})\rangle 
                  / \gamma_{\rm min}({\rm pair})}
       {m_{\rm p} / 2\gamma_{\rm min}({\rm nml})}
  \nonumber \\
  &=& 
  \frac{1}{9.18}
  \frac{\langle\gamma_-({\rm pair})\rangle}
       {\gamma_{\rm min}({\rm pair})}
  \frac{\gamma_{\rm min}({\rm nml})}
       {100}.
  \label{eq:LkinRat}
\end{eqnarray}
It follows from equation~(\ref{eq:def_gamAVR}) that
$\langle\gamma_-({\rm pair})\rangle / \gamma_{\rm min}({\rm pair}) \sim 3$
holds for $\alpha=-0.75$, for instance.
Thus, we obtain comparable kinetic luminosities irrespectively of
the composition assumed, 
if we evaluate them by the method described 
in \S~\ref{sec:Lkin_Core_Size}.

Let us examine the assumption of $K \sim 1$ in
equations~(\ref{eq:Lkin_3C345_1}), 
(\ref{eq:Lkin_3C279_1}), and (\ref{eq:Lkin_3C279_2}).
For a typical value of $\alpha= -0.75$, we obtain
$(\langle \gamma_- \rangle / \gamma_{\rm min})K
 = (\delta_{\rm eq}/\delta)^{17/2}$,
where $\delta_{\rm eq}$ refers to the 
\lq\lq equipartition Doppler factor'' defined by
(Readhead 1994)
\begin{eqnarray}
  \delta_{\rm eq}
  &\equiv& 
  \left\{ 10^{-3-6\alpha} F(\alpha)^{34} 
          \left[ \frac{0.65 h}
                      {q_0 z +(q_0-1)(-1+\sqrt{2q_0 z+1})}
          \right]^2
  \right.
  \nonumber \\
  &\times&
  \left.  (1+z)^{17-2\alpha}
          \nu_{\rm m}{}^{-35-2\alpha}
          S_{\rm m}{}^{16}
          \theta_{\rm d}{}^{-34}
  \right\}^{1/(13-2\alpha)},
  \label{eq:def_deleq}
\end{eqnarray}
where $F(\alpha)$ is given in Scott and Readhead (1977);
$\nu_{\rm m}$, $S_{\rm m}$, and $\theta_{\rm d}$ are measured in 
GHz, Jy, and mas, respectively.
Therefore, we can expect a rough energy equipartition 
between the radiating particles and the magnetic field
if $\delta_{\rm eq}$ is close the $\delta$ derived independently.

Substituting the input parameters presented in table~2,
we can compute $\delta_{\rm eq}$'s for the 3C~345 jet.
For component C5 at 1984.2, we obtain $\delta_{\rm eq}= 736$;
we thus discard this epoch for C5.
The computed $\delta_{\rm eq}$ ranges between 
$1.1$ and $67$, 
with the averaged value of 23 and
the standard deviation of 9.
The resultant magnetic field strength, $B =6.7 \pm 6.4$ mG, is reasonable.
In the same manner, for the 3C~279 jet,
we obtain $\delta=9.5 \pm 1.5$ and
$12 \pm 8$ mG for the four epochs.

Since $\delta_{\rm eq}= 23 \pm 9$ for 3C~345
is comparable with those presented in table~2
and since $\delta=9.5 \pm 1.5$ for 3C~279
is comparable with those adopted in \S~\ref{sec:3C279_individual}
(or table~3),
we can expect a rough energy equipartition in these two blazer jets.
However, it may be worth noting that the energy density 
of the radiating particles dominates that of the magnetic field
if $\delta$ becomes much smaller.
Such a case ($\delta= 4 \sim 12$) was discussed by 
Unwin et al. (1992; 1994; 1997),
assuming that a significant fraction of the X-rays 
was emitted via SSC from the radio-emitting components considered. 

We finally discuss the composition variation along the jet.
In the application to the two blazers, we assumed constant 
Lorentz factors ($\Gamma$) throughout the jet.
However, $\Gamma$ may in general decrease with increasing 
distance from the central engine.
If the decrease of $\Gamma$ is caused by an entrainment 
of the ambient matter (e.g., disk wind) consisting of a normal plasma, 
it follows from equation~(\ref{eq:Ne_nml}) that 
$\Nco({\rm nml})$ and hence $\Nco({\rm nml}) / \Nco({\rm SSA})$
increases at the place where the entrainment occurs.
(Note that the kinetic luminosity will not decrease due to an 
entrainment in a stationary jet.)
Therefore, the change of composition from a pair plasma into a normal
plasma along the jet may be found by the method described in this paper.

To study further the change of the composition, 
we need detailed spectral information 
($\nu_{\rm m}$, $S_{\rm m}$, $\alpha$)
of individual components along the jet.
To constrain the spectral turnover accurately, 
we must modelfit and decompose the VLBI images 
from high to low frequencies;
in another word, spatial resolution at low frequencies is crucial. 
Therefore, simultaneous observations with ground telescopes at higher
frequencies and with space+ground telescopes at lower frequencies
are preferable in this study.

\section*{Acknowledgments}

The author wishes to express his gratitude to
Drs. S. Kameno and A. P. Lobanov for valuable comments. 
He also thanks the Astronomical Data Analysis Center of
National Astronomical Observatory, Japan for the use of workstations.

\appendix
\section{Magnetic Field Strength Deduced from 
         Synchrotron Self-Absorption Thory}

The magnetic field strength, $B$, can be computed from 
the synchrotron spectrum, 
which is characterized by the peak frequency, $\nu_{\rm m}$, 
the peak flux density, $S_{\rm m}$,
and the spectral index, $\alpha$, in the optically thin regime.
Marscher (1983) considered a uniform spherical synchrotron source
and related $B$ with $\nu_{\rm m}$, $S_{\rm m}$, and $\alpha$
in the optically thin cases ($0 \geq \alpha \geq -1.0$).
Later, Cohen (1985) considered a uniform slab of plasma
as the synchrotron source and derived smaller values of $B$
for the same set of ($\nu$,$S_{\rm m}$,$\alpha$).
In this appendix, we express $B$ in terms of ($\nu$,$S_{\rm m}$,$\alpha$)
for a uniform sphere for arbitrary optical thickness
(i.e., $\alpha \leq 0$).
We assume that the magnetic field is uniform 
and that the distribution of radiating particles
are uniform in the configuration space 
and are isotropic and power-law (eq.[\ref{eq:ele-dist}]) 
in the momentum space.

\begin{figure} 
\centerline{ \epsfxsize=8cm \epsfbox[0 0 300 100]{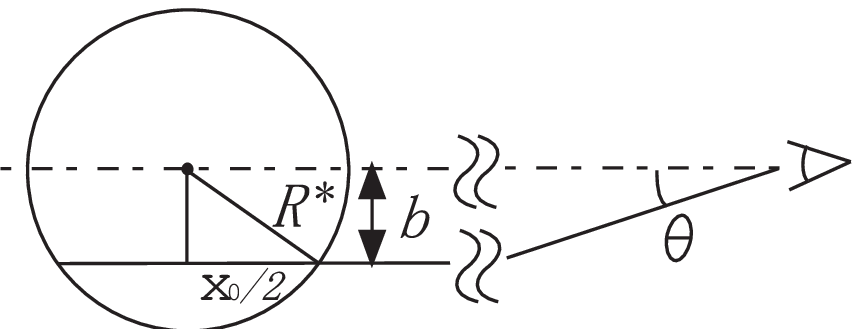} } 
\caption{\label{fig:thickness} 
Flux emitted from a spherical cloud of radius $R^\ast$.
        }
\end{figure}

For a uniform $B$ and $\Nco$, the transfer equation
gives the specific intensity
\begin{equation}
  I_\nu{}^\ast 
    = A \nu^\ast{}^{5/2}
      \left[ 1- \exp(-\alpha_\nu{}^\ast x_0{}^\ast) \right],
  \label{eq:intensity_1}
\end{equation}
where 
\begin{equation}
  A(\alpha)
  \equiv \left( \frac32 \right)^{-\alpha}
         \frac{e}{c}
         \frac{a(\alpha)}{C(\alpha)}
         \left( \frac{e}{2\pi m_{\rm e} c} \right)^{-3/2},
  \label{eq:def_A}
\end{equation}
and $x_0{}^\ast$ gives the physical thickness of the emitting region
along the line of sight.
At the distance $b$ away from the cloud center (fig.~\ref{fig:thickness}),
the fractional thickness becomes
\begin{equation}
  \frac{x_0{}^\ast}{2 R^\ast}
  = \sqrt{ 1 -\frac{b^2}{R^\ast{}^2} }
  = \sqrt{ 1 -\frac{4\theta^2}{\theta_{\rm d}^2} },
  \label{eq:frac_thick}
\end{equation}
which is Lorentz invariant.
Thus, the specific intensity at $b$,
or equivalently at angle $\theta$ away from the center, becomes
\begin{equation}
  I_\nu(\theta)
  = \left( \frac{\delta}{1+z} \right)^{1/2}
    A \nu^{5/2}
    \left[ 1-\exp\left( -\tau_\nu(0)\sqrt{1-\frac{4\theta^2}
                                                 {\theta_{\rm d}^2}}\,
                 \right)
    \right],
  \label{eq:intensity_2}
\end{equation}
where $\tau_\nu(0) \equiv \alpha_\nu{}^\ast \cdot 2R^\ast$ is the
optical depth for $\theta=0$ (or $b=0$);
we utilized $I_\nu / \nu^3 = I_\nu{}^\ast / \nu^\ast{}^3$ and
equation~(\ref{eq:LI-3}).
Even if special relativistic effects are important,
the observer find the shape to be circular.
We thus integrate $I_\nu(\theta)\cos\theta$ over 
the emitting solid angles 
$2\pi \sin\theta d\theta$ in $0 \leq \theta \leq \theta_{\rm d}/2$
to obtain the flux density
\begin{eqnarray}
  S_\nu &=& 2\pi \int_0^{\theta_{\rm d}/2}
                   I_\nu(\theta) \cos\theta\sin\theta d\theta
  \nonumber \\
        & & \hspace*{-1.2 truecm}
          = \frac{\pi}{4} \left( \frac{\theta_{\rm d}}{\rm rad} \right)^2
            \left(\frac{\delta}{1+z}\right)^{1/2} A \nu^{5/2}
            \int_0^1 \left( 1-{\rm e}^{-\tau_\nu(0)\sqrt{1-\chi}}\right)
            d\chi,
  \nonumber \\
        & & \ 
  \label{eq:flux_density}
\end{eqnarray}
where $\chi \equiv 4\theta^2/\theta_{\rm d}{}^2$ and $\theta \ll 1$
is used.

Putting the derivative $dS_\nu/d\nu$ to be $0$, we obtain
the equation that relates $\tau_\nu(0)$ and $\alpha$
at the turnover frequency, $\nu_{\rm m}$,
\begin{eqnarray}
  &&
  \int_0^1 \left[ 1-{\rm e}^{-\tau_\nu(0)\sqrt{1-\chi}} \right] d\chi
  \nonumber\\
  &=& \left( 1-\frac{2}{5}\alpha \right)
      \int_0^1 \tau_\nu(0)\sqrt{1-\chi}
               {\rm e}^{-\tau_\nu(0)\sqrt{1-\chi}} d\chi.
  \label{eq:opt_depth}  
\end{eqnarray}
We denote the solution $\tau_\nu(0)$ at $\nu=\nu_{\rm m}$ as
$\tau_{\rm m}(0)$,
which is presented as a function of $\alpha$ in table~1.

It is worth comparing equation~(\ref{eq:flux_density}) with 
that for a uniform slab of plasma with physical thickness $R^\ast$.
If the slab extends over a solid angle 
$\pi(\theta_{\rm d}/{\rm rad})^2/4$,
the flux density becomes
\begin{equation}
  S_\nu = \frac{\pi}{4} \left( \frac{\theta_{\rm d}}{\rm rad} \right)^2
          \left(\frac{\delta}{1+z}\right)^{1/2} A \nu^{5/2}
          \left( 1-{\rm e}^{-\tau_\nu(0)} \right).
  \label{eq:flux_density_slab}
\end{equation}
The optical depth $\tau_\nu(0)= \alpha_\nu^\ast \cdot 2R^\ast$
does not depend on $b$ (or $\theta$) for the slab geometry. 
Therefore, comparing equations~(\ref{eq:flux_density}) and 
(\ref{eq:flux_density_slab},)
we can define the effective optical depth, $\langle \tau_\nu \rangle$,
for a uniform sphere of plasma as
\begin{equation}
  {\rm e}^{-\langle \tau_\nu \rangle}
  \equiv \int_0^1 {\rm e}^{-\tau_\nu(0)\sqrt{1-\chi}} d\chi.
  \label{eq:mean_opt_depth}
\end{equation}
We denote $\langle\tau_\nu\rangle$ at $\nu=\nu_{\rm m}$ as
$\langle\tau_{\rm m}\rangle$,
which is presented in table~1.
Note that $\langle\tau_{\rm m}\rangle$ is less than 
$\tau_{\rm m}(0)$ due to the geometrical factor $\sqrt{1-\chi}$.

Using $\tau_{\rm m}(0)$, we can evaluate the peak flux density,
$S_{\rm m}$, at $\nu=\nu_{\rm m}$ 
and inversely solve equation~(\ref{eq:flux_density}) for $B$.
We thus obtain equation~(\ref{eq:lineA}) with
\begin{eqnarray}
  b(\alpha)
  &=& 3.98 \times 10^3 \left( \frac32 \right)^{-2\alpha}
      \left[ \frac{a(\alpha)}{C(\alpha)} \right]^2
  \nonumber\\
  \qquad 
  &\times& 
  \left\{ \int_0^1 \left[ 1-{\rm e}^{-\tau_{\rm m}(0)\sqrt{1-\chi}}
                   \right] d\chi
  \right\}^2.
  \label{eq:def_b}
\end{eqnarray}
The values are tabulated in table~1;
they are, in fact, close to the values obtained by Cohen (1985),
who presented $b(\alpha)$ for a slab geometry with $-\alpha \leq 1.25$.
This is because the averaged optical depth,
$\langle\tau_\nu\rangle$ at $\nu=\nu_{\rm m}$,
for a spherical geometry
becomes comparable with that for a slab geometry 
(Scott \& Readhead 1977). 

If we took the optically thin limit,
$\tau_{\rm m}(0) \ll 1$,
we could expand the integrant in equation~(\ref{eq:flux_density}) as
\begin{equation}
  1-{\rm e}^{-\tau_{\rm m}(0)\sqrt{1-\chi}}
  \approx \tau_{\rm m}(0)\sqrt{1-\chi},
  \label{eq:opt_depth_2}
\end{equation}
to obtain $b(-0.5)= 3.33$ and $b(-1.0)= 3.83$, for example.
This case ($\tau_{\rm m}\ll 1$) has been considered 
in many papers
for smaller $-\alpha$ ($\leq 1.0$).


\label{lastpage}
\end{document} 

\label{lastpage}

\end{document} 

\bye